\documentclass[twocolumn,prl,showpacs,amsmath,amssymb,superscriptaddress]{revtex4}

\usepackage{graphicx}
\usepackage{dcolumn}
\usepackage{bm}
\usepackage{amssymb}

\begin{document}
\title{Landau Cooling in Metal-Semiconductor Nanostructures}

\author{Francesco Giazotto}
\email{giazotto@sns.it}
\affiliation{NEST CNR-INFM and Scuola Normale Superiore, I-56126 Pisa, Italy}
\author{Fabio Taddei}
\affiliation{NEST CNR-INFM and Scuola Normale Superiore, I-56126 Pisa, Italy}\author{Michele Governale}
\affiliation{Institut f\"ur Theoretische Physik III, Ruhr-Universit\"at Bochum, D-44780 Bochum, Germany}
\author{Rosario Fazio}
\affiliation{International School for Advanced Studies (SISSA), via Beirut 2-4 I-34014, Trieste, Italy}
\affiliation{NEST CNR-INFM and Scuola Normale Superiore, I-56126 Pisa, Italy}
\author{Fabio Beltram}
\affiliation{NEST CNR-INFM and Scuola Normale Superiore, I-56126 Pisa, Italy}


\begin{abstract}
An electron-cooling principle based on Landau quantization is proposed for nanoscale conductor systems. Operation relies on energy-selective electron tunneling into a two-dimensional electron gas in quantizing magnetic fields.
This quantum refrigerator provides significant cooling power ($\sim10^{-9}$ W at a few K for realistic parameters) and offers a unique flexibility thanks to its tunability via the magnetic-field intensity.  The available performance is only marginally affected by nonidealities such as  disorder or imperfections in the semiconductor.
Methods for the implementation of this system and its characterization are discussed.     \end{abstract}


\maketitle

The relevance of understanding and tailoring thermal transport in nanoscale and mesoscopic conductors is becoming more and more apparent~\cite{giazottoRMP,diventra,pekolanature,quantthermcond}. This stems from the ever increasing miniaturization of the solid-state systems of interest for optoelectronics and advanced condensed-matter research which in turn demands a profound understanding of the generation and manipulation of heat flow. In this context electron cooling is probably one of the most relevant aspects.
Any solid-state system characterized by a gap in the density of states (DOS) at the Fermi energy is, in principle, suitable for \emph{refrigeration} schemes \cite{giazottoRMP}. In fact a gap provides an energy-selective electron-filtering mechanism which only allows \emph{hotter} quasiparticles to be extracted efficiently.
Superconductors represent prototype systems for refrigeration schemes operating around or below 1 K \cite{giazottoRMP,supercool}, and have yielded substantial electron temperature reduction in metallic nanostructures \cite{superelectron}.

Here we propose a novel mechanism for refrigeration based on \emph{Landau-level quantization}.
To this end we address thermal transport in two-dimensional electron gases (2DEGs)
showing that heat can be efficiently extracted from a metallic nanostructure, thanks to
the peculiar \emph{bulk} DOS originating from Landau quantization.
We find a large heat current through the system ($\sim10^{-9}$ W) yielding a sizable electron cooling also at bath temperatures as high as a few K. Moreover, the possibility to easily set the magnetic field value makes it possible to finely tune and optimize the performance available with this refrigeration principle under different operational conditions.

\begin{figure}[b!]
\includegraphics[width=8cm,clip]{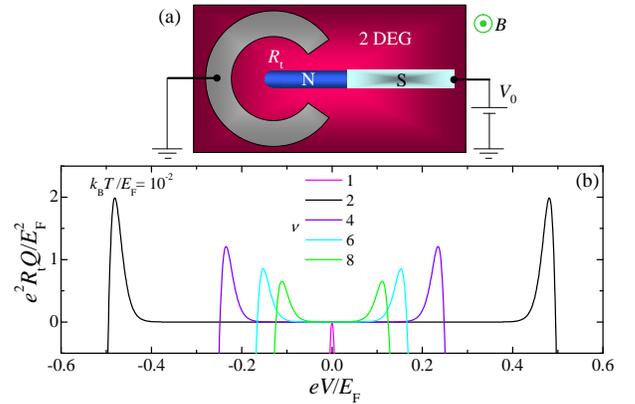}
\caption{(color) (a) A normal metal island (N) is tunnel coupled to a two-dimensional electron gas (2DEG) in an open pseudo Corbino disk geometry in the presence of a quantizing magnetic field. Gray region represents an ohmic contact to the 2DEG. 
(b) Heat current $Q$ extracted from the N region vs voltage drop $V$ across the barrier, calculated for different filling factors at $k_{\text{B}}T=10^{-2}E_{\text{F}}$. Other parameters used in the calculation \cite{parameters1}: $\tau E_{\text{F}}/(2\pi\hbar)=484$ and $\omega_{\text{Z}}/
\omega_{\text{c}}=14.7\times 10^{-3}$. }
\label{fig1}
\end{figure}

The system we envision is sketched in Fig. 1(a) and consists of a normal-metal island (N) coupled to the \emph{bulk} of a wide 2DEG region through a tunnel barrier of resistance $R_{\text{t}}$, in an open pseudo Corbino-disk geometry \cite{corbino,corbino1}.
A magnetic field $B$ is applied perpendicularly to the plane of the 2DEG, and the whole system is biased at voltage $V_0$ through a superconducting lead (S) in good metallic contact with N to prevent heat injection \cite{giazottoRMP}. In a quantizing magnetic field the bulk DOS $\mathcal{D}(E)$ of a 2DEG in the presence of impurities or disorder can be expressed as \cite{ando}
\begin{equation}
 \mathcal{D}(E)=\mathcal{C}\sum_{\sigma}\sum_{n=0}^{\infty}(\frac{\pi}{2}\Gamma^2)^{-\frac{1}{2}}
\,e^{-2\frac{(E+E_{\text{F}}-E_n^{\sigma})^2}{\Gamma^2}},
\label{DOS}
\end{equation}
where $E$ is the energy measured from the Fermi energy $E_{\text{F}}$, $\mathcal{C}=1/(2\pi \mathcal{D}_0 \ell _{\text{B}}^2)$,
$\mathcal{D}_0=m^*/(2 \pi \hbar^2)$ is the DOS per spin for $B=0$,
$\ell_{\text{B}}=\sqrt{\hbar/|eB|}$ denotes the magnetic length, $\Gamma=\sqrt{2\hbar^2\omega_{\text{c}}/(\pi\tau)}$, and 
$E_n^{\sigma}=(n+\frac{1}{2})\hbar\omega_{\text{c}}-\frac{1}{2}\sigma \hbar \omega_{\text{Z}}$.
 Furthermore, 
$\omega_{\text{c}}=eB/m^*$ is the cyclotron frequency, $\hbar\omega_{\text{Z}}=g^*\mu_{\text{B}}B$ is the Zeeman splitting, $m^*$ and $g^*$ are the electron effective mass and the effective \emph{g}-factor, respectively, $\tau$ is the relaxation time for $B=0$, $\mu_{\text{B}}$ is the Bohr magneton, and $\sigma =\pm 1$ is the spin index.
As we shall demonstrate below, the existence of Landau gaps will lead to a significant heat flow through the system thus implying \emph{hot}-excitation removal (i.e., \emph{refrigeration}) from the N island. 
Within a phenomenological tunneling Hamiltonian approach, the heat current $Q(V)$ flowing out the N side of the 2DEG/N junction can be written as \cite{giazottoRMP}
\begin{equation}
Q(V)=\frac{1}{e^2R_{\text{t}}}\int^{\infty}_{-\infty}dE\tilde{E}\mathcal{D}(E)[f_{\text{N}}(\tilde{E})-f_{\text{2DEG}}(E)],
\label{heatcurrent}
\end{equation}
where $\tilde{E}=E-eV$. $V$ is the voltage drop across the barrier, and $f_{\text{N(2DEG)}}$ is the Fermi distribution at temperature $T$ in the N island(2DEG). Furthermore, $V=V_0-V_{\text{th}}$ where $V_{\text{th}}$ is a threshold voltage \cite{vth}.
The heat current Eq. (\ref{heatcurrent}) can be rewritten as a series expansion in powers of $\Gamma/(k_{\text{B}}T)$:
\begin{eqnarray}
\nonumber Q(V)&=\frac{\mathcal{C}}{ e^2 R_{\text{t}}} \sum_{\sigma} \sum_{n=0}^{\infty} \sum_{m=0}^{\infty}
A_m^{\sigma} \left(\frac{\Gamma}{2}\right)^m \times\\& \left [f^{(m)}_{\text{N}}(E_n^{\sigma}-E_{\text{F}}-eV)-f^{(m)}_{\text{2DEG}}(E_n^{\sigma}-E_{\text{F}})\right]
\label{HCseries}
\end{eqnarray}
where $A_m^{\sigma}=\frac{1}{m!!}(E_n^{\sigma}-E_{\text{F}}-eV)$ if $m$ is even, $A_m^{\sigma}=\frac{1}{(m-1)!!}\frac{\Gamma}{2}$ if $m$ is odd, and $f^{(m)}_{\text{N(2DEG)}}=\partial^mf_{\text{N(2DEG)}}/\partial E^m$.
Equation (\ref{HCseries}) simplifies considerably the calculation of $Q(V)$ in the high-temperature limit, where only a few terms need to be retained.

Figure 1(b) shows the heat current $Q(V)$ versus bias voltage $V$ at $k_{\text{B}}T=10^{-2}E_{\text{F}}$ calculated for several  $\nu$ values \cite{parameters1}.
$Q$ is a symmetric function of $V$. For even $\nu$ it presents a \emph{positive} peak (describing heat extraction from the N region) in correspondence of the Landau level closest to the Fermi energy, and thereafter rapidly decreases to negative values. Furthermore, the maximum value of $Q$ decreases by increasing the filling factor, reflecting the DOS-amplitude reduction at smaller magnetic fields. For odd values of $\nu$ [see the case of $\nu=1$ in Fig. 1(b)] the heat current is always negative, unless Zeeman coupling completely lifts spin degeneracy giving rise to a gap \cite{gstar}. 	This full tunability of the heat current is a very attractive feature readily accessible by setting the magnitude of the applied magnetic field.  
The impact of disorder in the 2DEG on heat transport is analyzed in Fig. 2 (a). $Q(V)$ at $k_{\text{B}}T=10^{-2}E_{\text{F}}$ and $\nu=4$ is calculated for several  values of the relaxation time $\tau$. The latter is related to the 2DEG mobility, $\mu$, according to $\mu=e\tau/m^*$, and describes (with the chosen  parameters) a realistic GaAs 2DEG with mobility values in the range between $\sim 27$ and $1300$ m$^2$/Vs. In particular, by shortening the relaxation time (i.e., by increasing disorder) the positive $Q$ peak broadens and its amplitude is suppressed with the maximum moving towards lower bias voltage. This stems from disorder-induced broadening of the DOS [see Eq. (\ref{DOS})], and indicates that high-mobility 2DEGs are required to maximize $Q$. We stress, however, that a reduction of $\tau$ by a factor of $5$ (e.g., from $\tau = 5\times10^{-10}$ s to $1\times10^{-10}$ s for a GaAs 2DEG) yields only a small suppression of $Q$  ($\sim14\%$), and
that long relaxation times in 2DEGs can be achieved nowadays with epitaxial growth techniques: relaxation times as long as $2\times10^{-10}$ s (which corresponds to $\mu\sim 500$ m$^2$/Vs) and even larger can be obtained in the GaAs/AlGaAs material system \cite{roddaro1,roddaro2}.  \begin{figure}[t!]
\includegraphics[width=\columnwidth,clip]{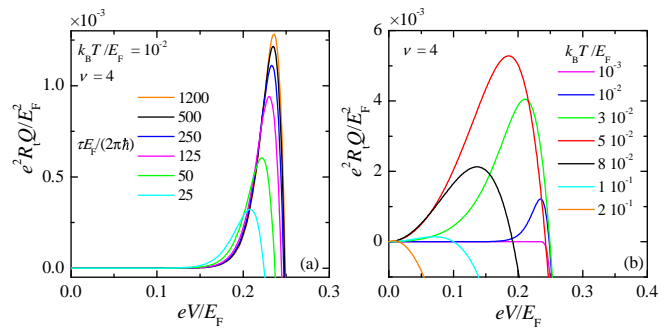}
\caption{(color) (a) Heat current $Q$ vs  $V$ calculated for several scattering times $\tau$ at $\nu =4$ and $k_{\text{B}}T=10^{-2}E_{\text{F}}$.  (b) $Q$ vs $V$ at $\nu =4$ calculated for different temperatures and  $\tau E_{\text{F}}/(2\pi\hbar)=484$. The Zeeman splitting is $\omega_{\text{Z}}/
\omega_{\text{c}}=14.7\times 10^{-3}$. }
\label{tS}
\end{figure}

The role of temperature on thermal transport is displayed in panel (b) of Fig. 2 where $Q$ vs $V$ at $\nu=4$ is shown for several values of $T$. At each temperature $Q$ is maximized at an optimal bias voltage ($V_{\text{opt}}$) which, at the lowest temperatures, approaches the separation between the Fermi energy and the closest Landau level. By increasing $T$, the heat-current amplitude grows and reaches its maximum $Q\simeq 5\times10^{-3}E_{\text{F}}^2/(e^2R_{\text{t}})$ at $V_{\text{opt}}\simeq0.185E_{\text{F}}/e$ for $k_{\text{B}}T=5\times10^{-2}E_{\text{F}}$.  A similar qualitative behavior also occurs at the other filling factors. At higher temperatures $Q$ becomes negative. 
The behavior of the maximum heat current $Q_{\text{max}}$ (i.e., that at the optimal bias voltage) as a function of temperature is shown in Fig. 3(a) for different filling factors $\nu$. In particular, at  low $T$, $Q_{\text{max}}$ is a bell-shaped function which, depending on $\nu$, is maximized at different temperatures. 
The optimal temperature decreases by increasing the filling factor while, at the same time, the window of positive heat current is reduced. For instance, at $\nu=2$ the heat current is maximized at $k_{\text{B}}T\simeq 0.1E_{\text{F}}$ where it reaches the value $Q_{\text{max}}^{\nu=2}\simeq 2.1\times 10^{-2}E_{\text{F}}^2/(e^2R_{\text{t}})$ while, at $\nu=4$, $Q_{\text{max}}^{\nu=4}\simeq 5.3\times 10^{-3}E_{\text{F}}^2/(e^2R_{\text{t}})$ at $k_{\text{B}}T\simeq 0.045E_{\text{F}}$. By choosing $R_{\text{t}}=500$ $\Omega$ the maximum heat current corresponds to $Q_{\text{max}}^{\nu=2}\simeq 4$ nW and $Q_{\text{max}}^{\nu=4}\simeq 1$ nW at $T\simeq 11.6$ K and $T\simeq 5.2$ K, respectively \cite{parameters1}.
Furthermore, for $k_{\text{B}}T\gtrsim 0.2E_{\text{F}}$, $Q_{\text{max}}(T)$ increases. This stems from the contribution to the heat current brought by additional Landau levels that occurs at higher temperatures.
\begin{figure}[t!]
\includegraphics[width=\columnwidth,clip]{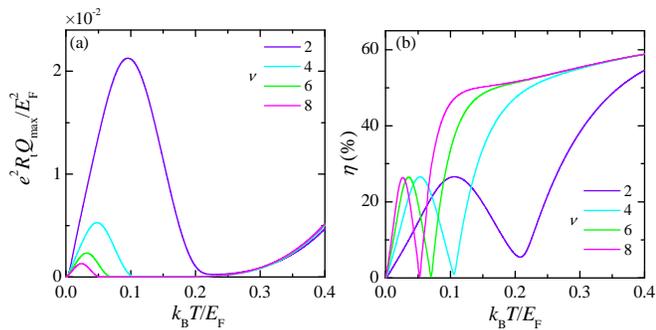}
\caption{(color) (a) $Q_{\text{max}}$ calculated at $V_{\text{opt}}$  vs $T$ for various filling factors. In the low temperature regime the heat current is maximized at different temperatures depending on $\nu$. (b) Efficiency $\eta$ vs $T$ at the same $\nu$ as in (a). Other parameters used in the calculation: $\tau E_{\text{F}}/(2\pi\hbar)=484$ and $\omega_{\text{Z}}/
\omega_{\text{c}}=14.7\times 10^{-3}$.}
\label{tF}
\end{figure}

Whether a cooling principle is usable in practice depends, in general, on its thermodynamic efficiency ($\eta$), defined as the ratio of heat extracted from the system and the work done by the source, $\eta(V)=Q(V)/[VI(V)]$ \cite{giazottoRMP}, where $I(V)=\frac{1}{eR_{\text{t}}}\int^{\infty}_{-\infty}dE\mathcal{D}(E)[f_{\text{N}}(\tilde{E})-f_{\text{2DEG}}(E)]$
is the charge current flowing through the 2DEG/N junction. 
The function $\eta$ versus $T$ calculated at $V_{\text{opt}}$ is shown in Fig. 3(b) for the same filling factors of Fig. 3(a). At low temperatures, $\eta$ is a non-monotonic function of $T$ and presents a relative maximum taking values as high as $\sim 27\%$ almost independently of $\nu$ (note that such a value for $\eta$ is comparable to the efficiency of a static superconductor-insulator-normal metal electron cooler which is of the order of $\sim25\%$ \cite{giazottoRMP}). After that, $\eta$ starts to increase monotonically at a temperature which depends on $\nu$, and saturates asymptotically around $60\%$ for large $T$. As we shall show in the following, however, only the lower-temperature behavior is relevant for electronic cooling. 

We now study the consequence of the positive heat current flowing off the normal metal, i.e.,  electronic refrigeration of the N island.
We focus on the quasi-equilibrium limit \cite{giazottoRMP} which can be attained at sufficiently low temperatures (i.e., for metallic systems typically around or below 1 K) when fast inelastic electron-electron relaxation drives the quasiparticle distribution toward a Fermi function at an effective temperature ($T_{\text{e}}$) which can differ significantly from that of the lattice ($T_{\text{bath}}$). This requires the tunneling rate of electrons in N to be (much) smaller than the electron-electron relaxation rate ($\tau_{\text{e-e}}^{-1}$). We note that in pure metals the latter is of the order of $\tau_{\text{e-e}}^{-1}\sim 10^8\ldots10^{10}$ s$^{-1}$ \cite{pierre} and even faster, so that the quasi-equilibrium regime turns out to be a realistic assumption for our system \cite{giazottoRMP}. In order to investigate the electronic refrigeration of the N island, we need to evaluate those mechanisms which tend to drive power into the N electrons. The main heat load comes from electron-phonon interaction with energy exchange given by \cite{roukes}
$Q_{\text{e-bath}}=\Sigma\mathcal{V}(T_{\text{e}}^5-T_{\text{bath}}^5)$,
where $\Sigma$ is the material-dependent electron-phonon coupling constant and $\mathcal{V}$ is the volume of the N region. Here, we neglect further sources of heat due to other relaxation mechanisms, for instance, radiative electron-photon heat load \cite{pekolanature,photons}, and we suppose negligible thermal-boundary (Kapitza) resistance between the N phonons and the 2DEG lattice \cite{giazottoRMP,rajauria}.
The stationary $T_{\text{e}}$ reached by the N island electrons is obtained by solving
\begin{equation}
Q(V,T_{\text{e}},T_{\text{bath}})+Q_{\text{e-bath}}(T_{\text{e}},T_{\text{bath}})=0,
\label{cool}
\end{equation}
which requires the sum of all thermal fluxes in the N island to be zero.
In writing Eq. (\ref{cool}) we set the electron temperature of the 2DEG equal to $T_{\text{bath}}$ \cite{giazottoRMP}, which is justified by our assumption of a macroscopic 2DEG reservoir where  electron-phonon interaction yields complete thermalization.
 
\begin{figure}[t!]
\includegraphics[width=\columnwidth,clip]{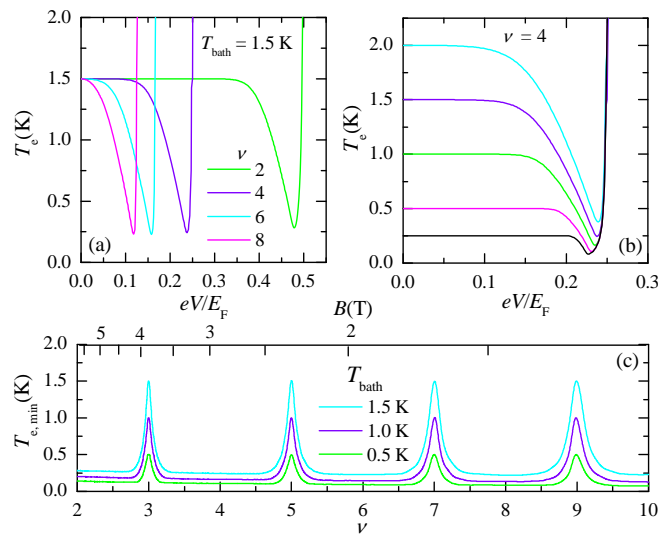}
\caption{ (color) (a) Temperature $T_{\text{e}}$ of the N island vs $V$  for different filling factors at  $T_{\text{bath}}=1.5$ K. (b) $T_{\text{e}}$ vs $V$ for $\nu =4$ at a few bath temperatures. (c) Minimum temperature $T_{\text{e,min}}$ of the N island vs $\nu$ at a few bath temperatures. The curves correspond to a realistic GaAs 2DEG with  $E_{\text{F}}=10$ meV, $\tau=2\times10^{-10}$ s, $\omega_{\text{Z}}/
\omega_{\text{c}}=14.7\times10^{-3}$,  and an aluminum island of volume $\mathcal{V}=5\times10^{-21}$ m$^3$, $\Sigma=2\times10^{8}$ WK$^{-5}$m$^{-3}$, and $R_{\text{t}}=500$ $\Omega$. }
\label{temp}
\end{figure}

In Fig. 4(a) we plot the steady-state electron temperature $T_{\text{e}}$ in N versus bias voltage $V$ at lattice temperature $T_{\text{bath}}=1.5$ K for a few filling factors. The electronic temperature is obtained as a solution of Eq. (\ref{cool}), assuming an aluminum (Al) island \cite{aluminum} of volume $\mathcal{V}=5\times10^{-21}$ m$^3$, $\Sigma=2\times10^{8}$ WK$^{-5}$m$^{-3}$ \cite{giazottoRMP}, and $R_{\text{t}}=500$ $\Omega$. For each $\nu$, the curves demonstrate a drastic reduction of $T_{\text{e}}$ with a minimum localized around the corresponding Landau level. Furthermore, the minimum of $T_{\text{e}}$ depends on $\nu$, i.e. on which Landau level is more effective at fixed $T_{\text{bath}}$ in removing hot excitations from the N island.
In particular, at $\nu=6$, a minimum $T_{\text{e}}$ of about $228$ mK can be achieved, which corresponds to temperature reductions of the order of $85\%$. We emphasize the magnitude of this value despite the large electron-phonon interaction present at such high lattice temperatures. Figure 4(b) displays $T_{\text{e}}$ versus $V$ at $\nu=4$ for different bath temperatures [$T_{\text{e}}(V=0)\equiv T_{\text{bath}}$]. The results show that a substantial electron temperature reduction can be achieved over a broad range of $T_{\text{bath}}$. At the lowest $T_{\text{bath}}$ the temperature reduction becomes less pronounced.
Nevertheless, the magnetic field can be tuned in order to optimize the cooling process even at the lowest bath temperatures.
The minimum achievable electron temperature ($T_{\text{e,min}}$) versus $\nu$ (or $B$) is shown in Fig. 4(c) for three different bath temperatures. $T_{\text{e,min}}$ shows an oscillatory behavior as a function of $\nu$ which can be understood as follows.
For odd $\nu$, no cooling is possible owing to the absence of a Landau gap at the Fermi energy. Consequently $T_{\text{e,min}}=T_{\text{bath}}$ giving rise to a maximum.
$T_{\text{e,min}}$ then takes its minimum values between two consecutive odd values of $\nu$. We also note the overall negative slope of the $T_{\text{e,min}}(\nu)$ curves in the regions between consecutive odd $\nu$ values. This fact indicates a reduction of the minimum achievable temperature at low $B$, which originates from the larger cooling efficiency provided by smaller Landau gaps at these bath temperatures [see Fig. 3(b)].

We should like to further comment on the experimental realization of this system. A 2DEG in the GaAs/AlGaAs material system is a suitable candidate to exhibit Landau quantization \cite{roddaro1,roddaro2}, although semiconductors with larger \emph{g}-factor (such as InAs 2DEGs) could provide additional flexibility thanks to enhanced magnetic-field-induced Zeeman splitting.   Furthermore, Fermi energy in the 2DEG can be changed through electrostatic gating, thus providing an additional degree of freedom for tuning the electronic temperature at different magnetic fields. As far as the S biasing contact is concerned [see Fig. 1(a)], a NbN electrode is particularly suitable to inject electric current in the N region. Its upper critical field can exceed 20 T at routinely accessible temperatures, thus allowing one to reach low $\nu$ at high magnetic fields without suppressing superconductivity \cite{akazaki}. In practice $T_{\text{e}}$ can be probed in the N island by Coulomb blockade thermometry \cite{CBT} which proved to be virtually immune to high magnetic fields \cite{field}. A shot-noise thermometer \cite{spietz} is a further possibility. Finally, we point out that the present system is to a large extent unaffected by nonidealities such as barrier nonuniformity which typically degrades cooling with a superconducting gap \cite{giazottoRMP}. 

In summary, we addressed thermal transport in metallic nanostructures contacted to the bulk of a 2DEG in quantizing magnetic fields showing that, thanks to the appearance of Landau gaps, a large heat current can flow off the N region. This leads to a significant electron refrigeration. The performance offered over a wide range of temperatures makes this principle attractive as a solid-state quantum refrigerator in nanoscale metal-semiconductor systems.    

We thank S. Roddaro, V. Pellegrini, V. Piazza, J. P. Pekola, and U. Z\"ulicke for helpful discussions. Partial financial support from HYSWITCH and NanoSci-ERA "NanoFridge" EU projects is acknowledged.


\begin{thebibliography}{99}
\bibitem{corbino}
Corbino geometry is the typical setup to probe the bulk states of a 2DEG in quantizing magnetic fields \cite{corbino1}.
\bibitem{parameters1}
The parameters used in the calculation correspond to a GaAs 2DEG with $m^*=0.067 m_{\text{e}}$, where $m_{\text{e}}$ is the free-electron mass, $g^*=0.44$, $\tau=2\times10^{-10}$ s, and $E_{\text{F}}=10$  meV \cite{roddaro1}.
\bibitem{gstar}
It is worthwhile to mention that the actual value of $g^*$  gets modified  in quantizing magnetic fields due to many-body effect.
Its exact value, however, is unimportant as far as this operational principle is concerned.
\bibitem{vth}
We note that $V_{\text{th}}$ originates from the insulating nature of the 2DEG within the first Landau gap. Its  value depends  on the actual geometry of the 2DEG/N system.  The physical description provided by Eq. (\ref{heatcurrent}) holds independently of its precise value, whose determination is beyond the scope of the present work.
\bibitem{aluminum}
Although superconducting at low temperature, Al will be driven into the normal state by the strong magnetic field required to induce Landau quantization.
\end{thebibliography}
\end{document}